\documentclass[aps,prl,twocolumn,superscriptaddress,floatfix]{revtex4-2}
\usepackage{graphicx,amsmath,amssymb,bm,xcolor}
\usepackage[colorlinks=true,linkcolor=blue,citecolor=blue,urlcolor=blue]{hyperref}
\newcommand{\Nf}{N_{\!f}}
\newcommand{\SM}{Supplemental Material~\cite{SM}}
\begin{document}

\title{Two quantum skyrmions form a magnon Josephson junction}

\author{Ricardo L. Silva}
\email{rmedasilva@gmail.com}
\affiliation{Departamento de Ci\^encias Naturais, Universidade Federal do Esp\'irito Santo, S\~ao Mateus, ES, Brazil}

\date{\today}

\begin{abstract}
Two quantum skyrmions in a frustrated spin-$1/2$ magnet bind into a molecule, and the molecule is a Josephson junction for magnons: the relative helicity $\theta$ and the number $n$ of magnons transferred between them obey $H=E_Cn^2-W\cos\theta$. DMRG on the $J_1$--$J_2$ triangular lattice gives $E_C$ as the curvature of the skyrmion energy at its natural size of seven flipped spins, and the measured helicity coherence fixes $W\simeq0.02J_1$, half the skyrmion--antiskyrmion tunnel splitting and independent of the anisotropy. Two helicity qubits side by side are thus strongly coupled by their own interaction.
\end{abstract}

\maketitle

\emph{Introduction.}---Magnetic skyrmions in centrosymmetric frustrated magnets are stabilized by competing exchange rather than by spin-orbit coupling~\cite{Okubo2012,Leonov2015,Lin2016,Hayami2016,Kurumaji2019}. They can be a few lattice spacings wide and carry an internal degree of freedom that costs no energy, the helicity $\chi$, the angle between the in-plane spin component and the radial direction~\cite{Nagaosa2013,Fert2017}. In the spin-$1/2$ limit the skyrmion survives as a bound state of $\Nf$ flipped spins in a ferromagnetic background~\cite{Lohani2019,Sotnikov2021,Haller2022,Siegl2022,Joshi2023,Haller2024,Petrovic2025}, and because the Hamiltonian is invariant under spin rotations about $z$, $\chi$ becomes a quantum rotor conjugate to $\Nf$~\cite{Psaroudaki2022}. This zero mode is the basis of the skyrmion helicity qubit~\cite{Psaroudaki2021,Xia2023,Psaroudaki2023}. The two-qubit gates proposed so far couple skyrmions in different layers or through a bus of magnons, phonons, or a cavity~\cite{Xia2023,Zou2023,Pan2024}; the most natural coupling, the interaction between two skyrmions lying side by side in the same film, has been treated only classically, where it depends on the relative helicity, $U=V(r)+W(r)\cos(\chi_A-\chi_B)$~\cite{Leonov2015,Zhang2017,Capic2020}. Quantum skyrmions themselves have been studied one at a time.

\begin{figure}[t]
\centering\includegraphics[width=\columnwidth]{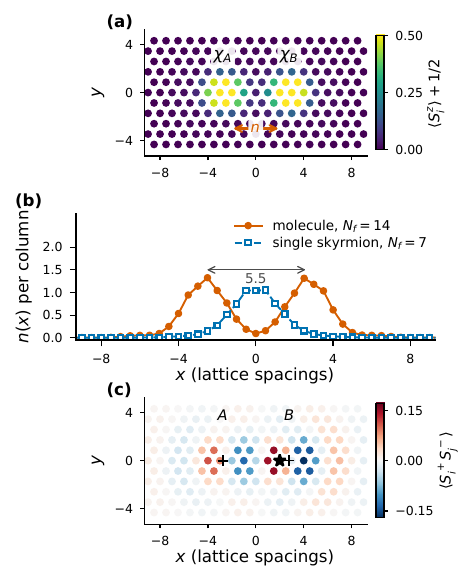}
\caption{The free skyrmion molecule ($K=0.03$, 261-site stadium, no pinning). (a) Flipped-spin density $\langle S^z_i\rangle+1/2$ in the $\Nf=14$ ground state: two skyrmions of seven flipped spins centered at $x=-2.75$ and $+2.79$. The number $n$ of magnons transferred between them is conjugate to the relative helicity $\theta=\chi_A-\chi_B$. (b) Flipped spins per column for the molecule and for one skyrmion in the same stadium. (c) In-plane correlation $\langle S^+_iS^-_j\rangle$ between every site $i$ and a wall site $j$ of skyrmion $B$ (star); crosses mark the lobe centers. The correlation reaches $0.16$ across the gap and follows the vortex pattern of skyrmion $A$.}
\label{fig:mol}
\end{figure}

Here we put two quantum skyrmions together (Fig.~\ref{fig:mol}). We find that they bind into a molecule and that the molecule is a Josephson junction for magnons: the relative helicity $\theta=\chi_A-\chi_B$ and the number $n$ of magnons transferred between the skyrmions are conjugate variables, the ``charging energy'' $E_Cn^2$ is set by the curvature of the skyrmion energy at its natural size, and the ``Josephson energy'' $-W\cos\theta$ is the quantized helicity-dependent interaction~\cite{Leggett2001,Smerzi1997,Albiez2005}. A one-parameter fit to the measured relative-helicity coherence accounts for most of the binding energy and is consistent with the measured magnon-number fluctuations, and it gives $W\simeq0.02J_1$, half the skyrmion-antiskyrmion tunnel splitting $\Delta_t$ of one skyrmion. Two helicity qubits placed side by side are thus strongly coupled by their own interaction.

\emph{Model and methods.}---We use the model of Ref.~\cite{Lohani2019} on the triangular lattice,
\begin{equation}
H=-J_1\!\sum_{\langle ij\rangle}\!\bm S_i\!\cdot\!\bm S_j+J_2\!\!\sum_{\langle\langle ij\rangle\rangle}\!\!\bm S_i\!\cdot\!\bm S_j-K\!\sum_{\langle ij\rangle}\!S_i^zS_j^z ,
\label{eq:H}
\end{equation}
with $J_1=1$, $J_2=0.5$ (frustration), and easy-axis anisotropy $K=0.02$--$0.04$. $S^z_{\rm tot}$ is conserved, so the number of flipped spins $\Nf$ labels sectors; a field is a constant in each sector and is omitted. Finite flakes, discs of 31--85 sites for one skyrmion and stadiums of half-width $r_0=3.7$--$5.5$ and up to 261 sites for two, are embedded in a frozen ferromagnetic background that enters as boundary fields. Flakes up to 31 sites are solved by exact diagonalization, which reproduces the selection rule $l_z=\pm(\Nf-1)\bmod6$ and the tunnel splitting $\Delta_t=0.050J_1$ ($K=0.05$) of Ref.~\cite{Lohani2019} and gives $\Delta_t=0.045J_1$ at $K=0.03$. Larger flakes are solved by DMRG~\cite{White1992,itensor} at fixed $\Nf$ with bond dimension $\chi$ up to 2000, validated against exact diagonalization. Sites are ordered column by column, so that the two halves of a stadium are contiguous and the distribution $P(n_A)$ of flipped spins in the left half follows from the charge blocks of the Schmidt decomposition. Methods, validation, and all convergence tests are given in the \SM.

\begin{figure}[t]
\centering\includegraphics[width=\columnwidth]{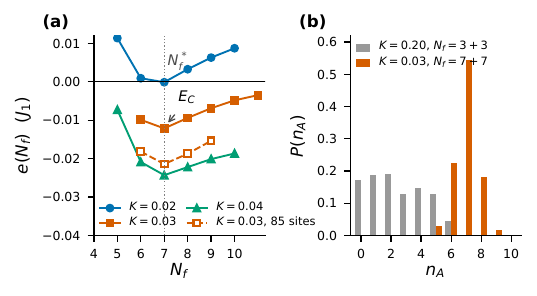}
\caption{(a) Binding energy per flipped spin of one quantum skyrmion on a 55-site disc for three anisotropies and on an 85-site disc for $K=0.03$. The minimum at $\Nf^*=7$ defines the natural size and its curvature the charging energy $E_C$. (b) Number of flipped spins in the left half of a stadium: for $K=0.20$ and $\Nf=3+3$ (gray) the two skyrmions have coalesced; for $K=0.03$ and $\Nf=7+7$ (color) the free molecule holds two skyrmions of seven.}
\label{fig:eNf}
\end{figure}

\emph{Natural size and coalescence.}---Let $e(\Nf)=[E(\Nf)-E_{\rm fm}-\Nf E^m_{\min}]/\Nf$ be the binding energy per flipped spin, with $E^m_{\min}=3K-1/4$ the bottom of the magnon band. Figure~\ref{fig:eNf}(a) shows that $e(\Nf)$ has a minimum at $\Nf^*=7$ for the three anisotropies and for discs of 55 and 85 sites: a quantum skyrmion has a natural size, set by the frustration, while $K$ sets only the depth [$e(7)=-0.0001$, $-0.012$, $-0.024$ for $K=0.02$, $0.03$, $0.04$ on the 55-site disc]. The curvature at the minimum,
\begin{equation}
E_C\equiv E(8)+E(6)-2E(7)=0.035\mbox{--}0.041\,J_1 ,
\label{eq:EC}
\end{equation}
is the energy cost of moving one flipped spin from one skyrmion of natural size to another; it will be the charging energy of the junction.

The minimum also decides whether two skyrmions can coexist. At fixed total $\Nf$ the Zeeman energy is constant, so two objects of $\Nf$ beat one object of $2\Nf$ only if $e(2\Nf)>e(\Nf)$. Where $e$ decreases monotonically, as at $K=0.20$ in the Ising-like regime, two small skyrmions placed at any distance coalesce: for $\Nf=3+3$ the interaction energy is $\simeq-1J_1$ and $P(n_A)$ has no peak at $n_A=3$ [Fig.~\ref{fig:eNf}(b)]; pinning wells only make it bimodal (\SM). Quantum skyrmions smaller than the natural size do not exist as pairs.

\emph{The molecule.}---Two skyrmions of natural size behave differently. With no pinning and no field, the $\Nf=14$ ground state on a 261-site stadium ($r_0=4.6$, $K=0.03$) consists of two lobes of $6.93$ and $7.07$ flipped spins separated by $5.5$ lattice spacings, with an almost empty gap between them and $P(n_A)$ peaked at 7 [Figs.~\ref{fig:mol} and \ref{fig:eNf}(b)]. With the single skyrmion computed at the center of the same stadium,
\begin{equation}
U_{\rm free}=E(14)-2E(7)=-0.0066\,J_1\quad(\chi=1200),
\label{eq:U}
\end{equation}
about $6\%$ of the binding energy of one skyrmion. The separation $5.5$ is the lowest-energy result of several DMRG starts, and it is the same at $K=0.02$, $0.03$, and $0.04$ and for stadiums of half-width $3.7$--$5.5$; the binding is $+0.002$, $-0.0066$, and $-0.006J_1$ at the three anisotropies, so the pair is unbound only at $K=0.02$, where the single skyrmion is itself marginal. The binding energy is quoted from the $r_0=4.6$ stadiums of length 12 and 16, in which neither the pair nor the reference skyrmion touches the rounded ends; its uncertainty is dominated by the bond-dimension convergence of the 14-spin state, $\pm0.002J_1$ (\SM). With weak pinning wells the interaction is repulsive below a separation of about 6 and attractive beyond, and penalty DMRG places the lowest excitations of the free molecule a few $10^{-3}J_1$ above the ground state, consistent with soft vibrational modes (\SM).

\emph{The magnon Josephson junction.}---In a fixed-$\Nf$ eigenstate the helicity of each skyrmion is undetermined, but the relative helicity is not: the operator $e^{i(\chi_A-\chi_B)}$ transfers one flipped spin from $B$ to $A$, so $\cos\theta$ is the magnon-tunneling operator between the two skyrmions. The classical interaction $W\cos(\chi_A-\chi_B)$ therefore becomes, for the relative coordinate of the molecule,
\begin{equation}
H_{\rm rel}=E_C\,n^2-W\cos\theta,\qquad [\theta,n]=i ,
\label{eq:JJ}
\end{equation}
the Hamiltonian of a Josephson junction~\cite{Leggett2001}, with $n$ the number of magnons transferred and $E_C$ the curvature of Eq.~(\ref{eq:EC}). Both sides of this picture are accessible to DMRG. (i) The coherence of the relative helicity. In the classical texture of a skyrmion of winding $\sigma$ the in-plane component on the wall is $S^+_i\propto e^{i(\sigma\varphi_i+\chi)}$, with $\varphi_i$ the polar angle of site $i$ around the center, so for equal windings $\langle S^+_iS^-_j\rangle\propto e^{i(\varphi_i-\varphi_j)}\langle e^{i\theta}\rangle$. Summing the wall-wall correlations (sites at $0.8$--$2.2$ from each lobe center) with the phases $e^{-i(\varphi_i-\varphi_j)}$ and normalizing by the same sums within each wall gives $|\langle e^{i\theta}\rangle|=0.441$ and $0.443$ on the 261- and 217-site stadiums at $K=0.03$, $0.416$ at $K=0.04$, and $0.471$ at $K=0.02$, against $\simeq0.20$ with opposite-winding phases and $\simeq0$ without phases; the individual correlations reach $0.16$ across the gap, out of a maximum of $0.25$ [Fig.~\ref{fig:mol}(c)]. Changing the wall window moves the value between $0.42$ and $0.53$. (ii) The magnon-number fluctuations. The numbers of flipped spins in the two skyrmions are anticorrelated, ${\rm cov}(N_A,N_B)=-0.45$ for discs of radius $2.6$ around the two lobes, and the variance of $N_A$ exceeds that of an isolated skyrmion in the same disc by $0.24$; across the midpoint cut the excess variance over two independent tails is $0.14$. These numbers depend on the partition, because any static boundary cuts through the skyrmion walls and the two tails exclude each other in the gap, so they bound $\langle n^2\rangle$ to $0.1$--$0.45$ rather than measure it. (iii) The number entropy of $P(n_A)$ exceeds that of two independent tails by $0.19$ nats, and the mutual information between the two halves exceeds the independent value by $\sim0.5$ nats (\SM).

\begin{figure}[t]
\centering\includegraphics[width=\columnwidth]{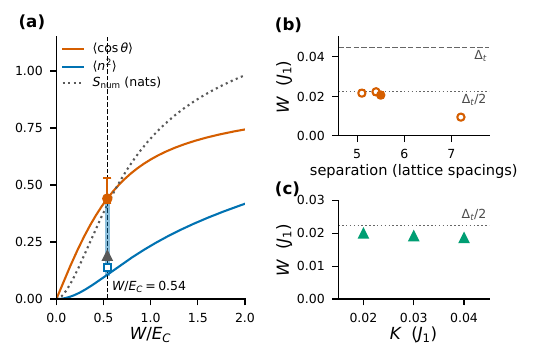}
\caption{(a) Ground state of the rotor of Eq.~(\ref{eq:JJ}): relative-helicity coherence $\langle\cos\theta\rangle$, magnon-number fluctuation $\langle n^2\rangle$, and number entropy $S_{\rm num}$ versus $W/E_C$. The measured coherence (circle; bar: range over wall windows) fixes $W/E_C=0.54$ (dashed line). At that ratio, the shaded bar is the partition-dependent range of the measured magnon-number fluctuation, the open square the excess variance across the midpoint cut, and the triangle the measured excess of number entropy, a lower bound on $S_{\rm num}$. (b) Josephson coupling $W$ versus the separation of the two skyrmions at $K=0.03$ (open circles, weakly pinned pairs; filled circle, free molecule); error bars span $E_C=0.035$--$0.041J_1$. (c) $W$ of the free molecule versus anisotropy. Dashed and dotted lines: $\Delta_t$ and $\Delta_t/2$ of a single skyrmion at $K=0.03$.}
\label{fig:JJ}
\end{figure}

Equation~(\ref{eq:JJ}) has one free ratio, $W/E_C$. Fixing it with $\langle\cos\theta\rangle=0.44$ gives $W/E_C=0.54$ [Fig.~\ref{fig:JJ}(a)], and the model then predicts a ground-state energy of $-0.0046$ to $-0.0054J_1$ relative to decoupled rotors, against the measured binding $-0.0066\pm0.002J_1$: most of the binding of the molecule is Josephson energy, and the two tail-free observables, coherence and energy, agree to about $25\%$ with a single parameter. The model also gives $\langle n^2\rangle=0.11$ and a number entropy of $0.41$ nats, inside the measured range of $\langle n^2\rangle$ and above the measured entropy excess, which is a lower bound on it; and it predicts that the number fluctuations fall by a factor $3.6$ between separations $5.4$ and $7.2$, where the measured covariance falls by $2.9$. With $E_C=0.035$--$0.041J_1$,
\begin{equation}
W=0.019\mbox{--}0.022\,J_1\simeq\tfrac12\Delta_t ,
\label{eq:W}
\end{equation}
and the same analysis at $K=0.02$ and $0.04$, with the charging energies of those anisotropies, gives $W=0.020$ and $0.019J_1$ [Fig.~\ref{fig:JJ}(c)]: like $\Nf^*$ and the equilibrium separation, the Josephson coupling is set by the frustration and not by the anisotropy, which changes only $E_C$. The coupling decays with distance. In weakly pinned pairs that settle at separations $5.1$, $5.4$, and $7.2$ the coherence is $0.46$, $0.46$, and $0.24$, that is, $W/E_C=0.57$, $0.58$, and $0.25$ and $W\simeq0.021$, $0.022$, and $0.009J_1$ [Fig.~\ref{fig:JJ}(b)]: the coupling halves over two lattice spacings.

\emph{Discussion.}---Three scales characterize the molecule: $E_C\simeq0.04J_1$, $W\simeq0.02J_1$, and $\Delta_t\simeq0.045J_1$. The ratio $W/E_C\simeq0.5$ places the junction in the crossover between the number-locked ($W\ll E_C$) and phase-locked ($W\gg E_C$) regimes, where the relative helicity is partially coherent and the number of transferred magnons fluctuates by less than one. Josephson physics of magnons has been discussed for magnon condensates~\cite{Nakata2014,Troncoso2014}; here the two reservoirs are single quantum skyrmions, the transferred particles are the flipped spins that constitute them, and the phase is their relative helicity. For the helicity-qubit program~\cite{Psaroudaki2021,Xia2023} the consequence is direct: two skyrmion qubits in the same layer do not need an engineered bus. Their own interaction is a Josephson coupling of order $\Delta_t/2$, and the molecule assembles itself at the separation where it acts; the coupling can be tuned through the distance, which pinning or a potential landscape can set.

Several points are left open. The spectrum of the four-state manifold $\{\sigma_A\sigma_B\}$ of skyrmion and antiskyrmion windings lies below the soft modes of the molecule and could not be resolved by penalty DMRG; the finite opposite-winding harmonic in the wall-wall correlations (\SM) shows that each constituent is, like the isolated skyrmion~\cite{Lohani2019,Liscak2026}, a superposition of both windings, and a symmetry-resolved method is needed to separate the manifold. The $20$--$30\%$ discrepancies between the rotor model and the DMRG data are of the size expected from the approximate subtraction of the skyrmion tails, from the wall-window dependence of the coherence, and from the harmonic truncation of the rotor, and could be reduced with a two-rotor model that includes $\Delta_t$. We have worked at $J_2/J_1=0.5$; at larger frustration the few-spin ground states carry different angular momenta and are not the same objects. The frustrated magnet with a few-spin skyrmion is, in any case, the regime where this physics lives: in the dipolar skyrmions of thin films the helicity is pinned to the Bloch value and $10^{4}$--$10^{5}$ spins suppress any helicity coherence. Candidate systems are the centrosymmetric frustrated magnets in which skyrmion lattices have been observed~\cite{Kurumaji2019} and van der Waals magnets with competing exchange.

\emph{Conclusion.}---A quantum skyrmion in a frustrated magnet has a natural size of $\Nf^*=7$ flipped spins; below it, skyrmions coalesce. Two skyrmions of natural size form a bound molecule that is a magnon Josephson junction, $H=E_Cn^2-W\cos\theta$, with $E_C$ the curvature of the skyrmion energy at $\Nf^*$ and $W\simeq0.02J_1\simeq\Delta_t/2$ the quantized helicity interaction, independent of the anisotropy. The two helicity qubits of the molecule are strongly coupled by their own interaction.

\begin{acknowledgments}
This work received no external funding. All calculations were carried out on the author's own equipment. The DMRG calculations used the ITensor library~\cite{itensor}. The data that support the findings of this Letter are openly available in the Zenodo repository~\cite{zenodo}.
\end{acknowledgments}

\bibliographystyle{apsrev4-2}
\bibliography{refs}
\end{document}